\documentclass{emulateapj}

\def\rgb{RGB~J0521.8+2112}
\def\ver{VER~J0521+211}
\def\lat{{\em Fermi}-LAT}
\usepackage{unites2e}
\usepackage{amssymb}
\usepackage{url}
\usepackage{natbib}

\begin{document}

\title{Discovery of a New TeV Gamma-Ray Source: VER~J0521+211}

\author{
S.~Archambault\altaffilmark{1},
T.~Arlen\altaffilmark{2},
T.~Aune\altaffilmark{2},
B.~Behera\altaffilmark{3},
M.~Beilicke\altaffilmark{4},
W.~Benbow\altaffilmark{5},
R.~Bird\altaffilmark{6},
A.~Bouvier\altaffilmark{7},
J.~H.~Buckley\altaffilmark{4},
V.~Bugaev\altaffilmark{4},
K.~Byrum\altaffilmark{8},
A.~Cesarini\altaffilmark{9},
L.~Ciupik\altaffilmark{10},
M.~P.~Connolly\altaffilmark{9},
W.~Cui\altaffilmark{11},
M.~Errando\altaffilmark{12},
A.~Falcone\altaffilmark{13},
S.~Federici\altaffilmark{3,14},
Q.~Feng\altaffilmark{11},
J.~P.~Finley\altaffilmark{11},
L.~Fortson\altaffilmark{15},
A.~Furniss\altaffilmark{7},
N.~Galante\altaffilmark{5},
D.~Gall\altaffilmark{16},
G.~H.~Gillanders\altaffilmark{9},
S.~Griffin\altaffilmark{1},
J.~Grube\altaffilmark{10},
G.~Gyuk\altaffilmark{10},
D.~Hanna\altaffilmark{1},
J.~Holder\altaffilmark{17},
G.~Hughes\altaffilmark{3},
T.~B.~Humensky\altaffilmark{18},
P.~Kaaret\altaffilmark{16},
M.~Kertzman\altaffilmark{19},
Y.~Khassen\altaffilmark{6},
D.~Kieda\altaffilmark{20},
H.~Krawczynski\altaffilmark{3},
F.~Krennrich\altaffilmark{21},
S.~Kumar\altaffilmark{17},
M.~J.~Lang\altaffilmark{9},
A.~S~Madhavan\altaffilmark{21},
G.~Maier\altaffilmark{3},
P.~Majumdar\altaffilmark{2,22},
S.~McArthur\altaffilmark{23},
A.~McCann\altaffilmark{24},
J.~Millis\altaffilmark{25},
P.~Moriarty\altaffilmark{26},
R.~Mukherjee\altaffilmark{12},
A.~O'Faol\'{a}in de Bhr\'{o}ithe\altaffilmark{6},
R.~A.~Ong\altaffilmark{2},
A.~N.~Otte\altaffilmark{27},
N.~Park\altaffilmark{23},
J.~S.~Perkins\altaffilmark{28},
M.~Pohl\altaffilmark{14,3},
A.~Popkow\altaffilmark{2},
H.~Prokoph\altaffilmark{3},
J.~Quinn\altaffilmark{6},
K.~Ragan\altaffilmark{1},
L.~C.~Reyes\altaffilmark{29},
P.~T.~Reynolds\altaffilmark{30},
G.~T.~Richards\altaffilmark{27},
E.~Roache\altaffilmark{5},
D.~B.~Saxon\altaffilmark{17},
G.~H.~Sembroski\altaffilmark{11},
A.~W.~Smith\altaffilmark{20},
D.~Staszak\altaffilmark{1},
I.~Telezhinsky\altaffilmark{14,3},
M.~Theiling\altaffilmark{11},
A.~Varlotta\altaffilmark{11},
V.~V.~Vassiliev\altaffilmark{2},
S.~Vincent\altaffilmark{3},
S.~P.~Wakely\altaffilmark{23},
T.~C.~Weekes\altaffilmark{5},
A.~Weinstein\altaffilmark{21},
R.~Welsing\altaffilmark{3},
D.~A.~Williams\altaffilmark{7},
B.~Zitzer\altaffilmark{8}\\
(The VERITAS Collaboration)\\
M.~B\"{o}ttcher\altaffilmark{31},
S.~J.~Fegan\altaffilmark{32},
P.~Fortin\altaffilmark{5},
J.~P.~Halpern\altaffilmark{33},
Y.~Y.~Kovalev\altaffilmark{34,35},
M.~L.~Lister\altaffilmark{36},
J.~Liu\altaffilmark{33},
A.~B.~Pushkarev\altaffilmark{37,38},
P.~S.~Smith\altaffilmark{39}
}
\altaffiltext{1}{Physics Department, McGill University, Montreal, QC H3A 2T8, Canada}
\altaffiltext{2}{Department of Physics and Astronomy, University of California, Los Angeles, CA 90095, USA}
\altaffiltext{3}{DESY, Platanenallee 6, 15738 Zeuthen, Germany}
\altaffiltext{4}{Department of Physics, Washington University, St. Louis, MO 63130, USA}
\altaffiltext{5}{Fred Lawrence Whipple Observatory, Harvard-Smithsonian Center for Astrophysics, 
Amado, AZ 85645, USA; \mbox{fortin@veritas.sao.arizona.edu}}
\altaffiltext{6}{School of Physics, University College Dublin, Belfield, Dublin 4, Ireland}
\altaffiltext{7}{Santa Cruz Institute for Particle Physics and Department of Physics, University of 
California, Santa Cruz, CA 95064, USA}
\altaffiltext{8}{Argonne National Laboratory, 9700 S. Cass Avenue, Argonne, IL 60439, USA}
\altaffiltext{9}{School of Physics, National University of Ireland Galway, University Road, Galway, 
Ireland}
\altaffiltext{10}{Astronomy Department, Adler Planetarium and Astronomy Museum, Chicago, IL 60605, 
USA}
\altaffiltext{11}{Department of Physics, Purdue University, West Lafayette, IN 47907, USA }
\altaffiltext{12}{Department of Physics and Astronomy, Barnard College, Columbia University, NY 
10027, USA; \mbox{errando@astro.columbia.edu}}
\altaffiltext{13}{Department of Astronomy and Astrophysics, 525 Davey Lab, Pennsylvania State 
University, University Park, PA 16802, USA}
\altaffiltext{14}{Institute of Physics and Astronomy, University of Potsdam, 14476 Potsdam-Golm, 
Germany}
\altaffiltext{15}{School of Physics and Astronomy, University of Minnesota, Minneapolis, MN 55455, USA}
\altaffiltext{16}{Department of Physics and Astronomy, University of Iowa, Van Allen Hall, Iowa City, IA 52242, USA}
\altaffiltext{17}{Department of Physics and Astronomy and the Bartol Research Institute, University of Delaware, Newark, DE 19716, USA; \mbox{jholder@physics.udel.edu}}
\altaffiltext{18}{Physics Department, Columbia University, New York, NY 10027, USA}
\altaffiltext{19}{Department of Physics and Astronomy, DePauw University, Greencastle, IN 46135-0037, USA}
\altaffiltext{20}{Department of Physics and Astronomy, University of Utah, Salt Lake City, UT 84112, USA}
\altaffiltext{21}{Department of Physics and Astronomy, Iowa State University, Ames, IA 50011, USA}
\altaffiltext{22}{Saha Institute of Nuclear Physics, Kolkata 700064, India}
\altaffiltext{23}{Enrico Fermi Institute, University of Chicago, Chicago, IL 60637, USA}
\altaffiltext{24}{Kavli Institute for Cosmological Physics, University of Chicago, Chicago, IL 60637, USA}
\altaffiltext{25}{Department of Physics, Anderson University, 1100 East 5th Street, Anderson, IN 46012}
\altaffiltext{26}{Department of Life and Physical Sciences, Galway-Mayo Institute of Technology, Dublin Road, Galway, Ireland}
\altaffiltext{27}{School of Physics and Center for Relativistic Astrophysics, Georgia Institute of Technology, 837 State Street NW, Atlanta, GA 30332-0430}
\altaffiltext{28}{N.A.S.A./Goddard Space-Flight Center, Code 661, Greenbelt, MD 20771, USA}
\altaffiltext{29}{Physics Department, California Polytechnic State University, San Luis Obispo, CA 
94307, USA}
\altaffiltext{30}{Department of Applied Physics and Instrumentation, Cork Institute of Technology, 
Bishopstown, Cork, Ireland}
\altaffiltext{31}{Centre for Space Research, North-West University, Potchefstroom, 2531, 
South Africa}
\altaffiltext{32}{Laboratoire Leprince-Ringuet, \'Ecole polytechnique, 
CNRS/IN2P3, Palaiseau, France; \mbox{sfegan@llr.in2p3.fr}}
\altaffiltext{33}{Columbia Astrophysics Laboratory, Columbia University, New York, NY 10027, USA}
\altaffiltext{34}{Max-Planck-Institut f\"{u}r Radioastronomie, Auf dem H\"{u}gel 69, D-53121 Bonn,
Germany}
\altaffiltext{35}{Astro Space Center of Lebedev Physical Institute, Profsoyuznaya Str. 84/32, 117997
Moscow, Russia}
\altaffiltext{36}{Department of Physics, Purdue University, 525 Northwestern Avenue, West Lafayette,
IN 47907, USA}
\altaffiltext{37}{Pulkovo Astronomical Observatory, Pulkovskoe Chaussee 65/1, 196140 St. Petersburg,
Russia}
\altaffiltext{38}{Crimean Astrophysical Observatory, 98409 Nauchny, Crimea, Ukraine}
\altaffiltext{39}{Steward Observatory, University of Arizona, Tucson, AZ 85721, USA}

\begin{abstract}

We report the detection of a new TeV gamma-ray source, \object[VER J0521+211]{VER~J0521+211}, based
on observations made with the VERITAS imaging atmospheric Cherenkov telescope array. 
These observations were motivated by the discovery of a cluster of $>30\U{GeV}$ photons in the 
first year of \lat\ observations. 
\ver\ is relatively bright at TeV energies, with a mean photon flux of $(1.93
\pm 0.13_{\mathrm{stat}} \pm 0.78_{\mathrm{sys}})\times 10^{-11}\UU{cm}{-2}\UU{s}{-1}$ above
$0.2\U{TeV}$ during the period of the VERITAS observations. The source is strongly variable on a
daily timescale across all wavebands, from optical to TeV, with a peak flux corresponding to 
$\sim 0.3$ times the steady Crab Nebula flux at TeV energies. Follow-up observations in the optical 
and X-ray bands classify the newly-discovered TeV source as a BL Lac-type blazar with uncertain
redshift, although recent measurements suggest $z=0.108$. \ver\ exhibits all the defining
properties of blazars in radio, optical, X-ray, and gamma-ray wavelengths. 
\end{abstract}


\section{Introduction}

\begin{figure*}[tbp]
\center
\includegraphics[width=0.7\textwidth]{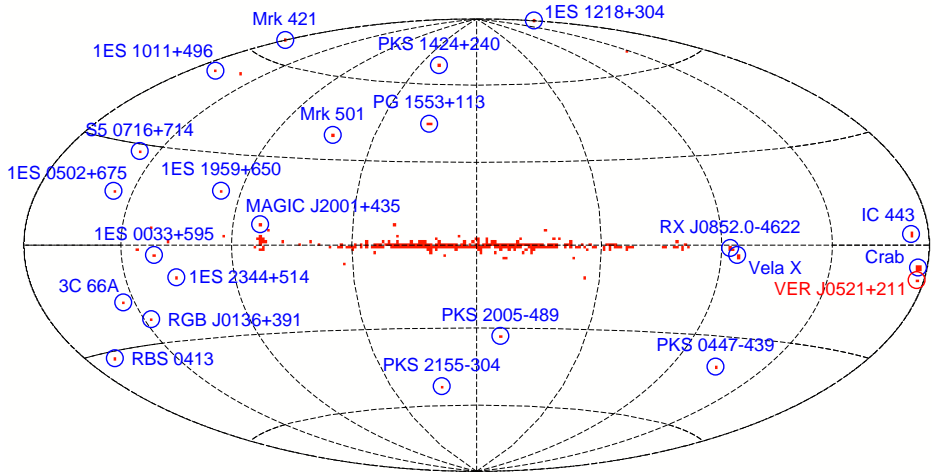} 
\caption{A map, in Galactic coordinates, showing the arrival
directions of all \textit{diffuse}-class photon events$^{41}$ with $zenith\ angle <105^{\circ}$ 
recorded by
\textit{Fermi}-LAT between 2008 Aug 4 and 2009 Aug 25 with reconstructed energies greater than
$30\U{GeV}$. The events are binned in $1^{\circ}$ bins, and bins containing at least 5 events
are shaded red. Known TeV sources coincident with shaded bins are indicated in regions where source 
confusion or diffuse background is not an issue.
\label{Fermi_skymap}}
\end{figure*}

TeV astronomy is concerned with the
detection of astrophysical gamma rays with energies greater than
$\sim 0.1\U{TeV}$. The most sensitive detectors in this energy region
are the current generation of ground-based imaging atmospheric Cherenkov telescopes: 
VERITAS \citep{VERITAS}, H.E.S.S. \citep{HESS} and MAGIC \citep{MAGIC},
which can detect sources with a flux less than 0.01 times the steady flux
of the Crab Nebula \citep[][hereafter ``Crab'']{hillas-crab} with an exposure of a few
tens of hours. The angular
resolution of current instruments
is $\sim 0.1^{\circ}$, and the field of view is
typically limited to a diameter of $<5^{\circ}$. Observations 
consist of surveying regions of the sky using many
overlapping, noncontemporaneous exposures \citep{HESS_survey,
VERITAS_survey}, or targeting locations of interest based on
information provided by observations at other wavelengths. 

The Large Area Telescope (LAT) aboard the \textit{Fermi Gamma-ray Space Telescope}
is the first instrument to provide a view of the entire gamma-ray sky at energies that overlap
with those accessible to ground-based telescopes.
Figure~\ref{Fermi_skymap} shows the arrival directions of all
\textit{diffuse}-class photon events\footnote{ IRF version \texttt{P6V3},
see {http://fermi.gsfc.nasa.gov/ssc/data/
analysis/documentation/Cicerone/Cicerone\_Data/LAT\_DP.html
} } recorded during the first year of
\textit{Fermi}-LAT observations with energies above $30\U{GeV}$. 
Clearly, the highest-energy
\textit{Fermi}-LAT observations provide an excellent guide to the TeV
sky, and can be used to select additional TeV candidates for targeted
observations. Numerous authors have addressed this, producing
catalogs based on $>100\U{GeV}$ \lat\ events
\citep{Neronov_exgal} and lists of TeV source candidates
\citep[e.g.,][]{Fegan_AGN}. 
In this work, the presence of a cluster of high-energy photons (Figure~\ref{Fermi_skymap}) 
spatially associated with an X-ray source was used to motivate
observations with the VERITAS array of a previously unobserved
location, leading to the detection of a new bright TeV source: \ver. The only other cluster flagged by this method (1ES~0502+675) was already being observed by VERITAS at the time and was 
subsequently detected after 13 hours of exposure \citep{0502}.

This paper presents the discovery of gamma-ray emission from \ver, and a detailed study of its
observational properties from radio frequencies to gamma-ray energies. The paper is structured as
follows: Sections~2 to 5 present the observations and analysis results by VERITAS (TeV), \lat\
(GeV), {\it Swift} (X-ray), and MDM and Steward Observatory (optical), respectively. The
identification of \ver\ as a new TeV blazar is discussed in Section~6, and its main observational
properties are presented in Section~7. An upper limit on the redshift of \ver\ is calculated in
Section~8, and the overall spectral energy distribution is discussed in Section~9. Finally,
Section~10 summarizes the main conclusions of the study.

\section{VERITAS observations}

The VERITAS observatory is described in detail in \citet{VERITAS} and
\citet{VERITAST1}. 
The array consists of four
12\,m-diameter imaging atmospheric Cherenkov telescopes, 
with photomultiplier (PMT) cameras 
covering a field of view of
$3.5^{\circ}$.
The array has a total effective area of $\sim 5 \times 10^4\UU{m}{2}$ between 0.2 and 10\U{TeV}.  
Following the relocation of the original prototype
telescope to a more favorable position in Summer 2009
\citep{T1_relocation}, VERITAS has 
sensitivity to detect a source with 0.01\U{Crab}
flux in under 30\U{hours} of observations. The angular
and energy resolution for reconstructed gamma-ray showers is energy
dependent, reaching $\sim 0.1 ^{\circ}$ and 15\%, respectively, at
$1\U{TeV}$.

Inspection of the $>30\U{GeV}$ \textit{Fermi}-LAT map
(Figure~\ref{Fermi_skymap}) led to the identification of a cluster of high-energy
photons, which was used to trigger VERITAS observations  
centered around 
R.A. = $05^{\rm h}21^{\rm m}46^{\rm s}$,
Decl. = $+21^{\circ}12^{\prime}51.\!^{\prime\prime}5$ (J2000),
corresponding to the position of \object[RGB J0521.8+2112]{RGB~J0521.8+2112}, the only radio/X-ray
source within $0.1^{\circ}$ of the
LAT excess. 

Observations took place between MJD 55126 (2009 Oct 22)
and MJD 55212 (2010 Jan 16) and consisted of 20-minute exposures
in \textit{wobble} observing mode \citep{fomin}, taken at a mean zenith angle of $16^{\circ}$.
After excluding data taken under poor weather
conditions or with hardware problems, the dataset comprises 14.5
hours livetime.

 \begin{figure}[tdp]
 \center
 \includegraphics[width=0.95\columnwidth]{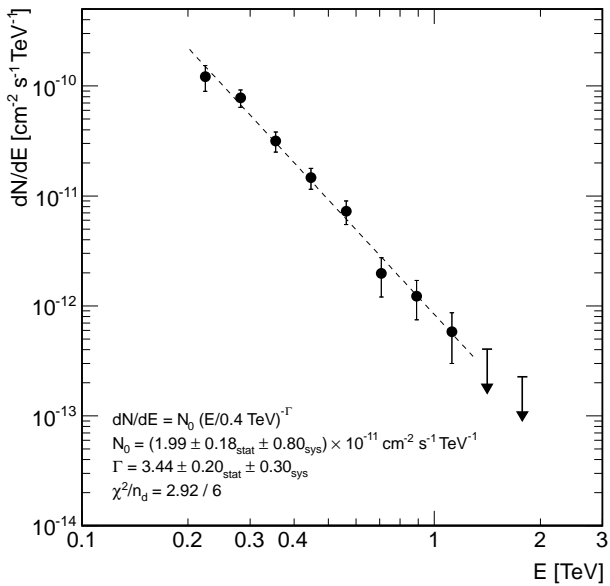} 
 \caption{Time-averaged differential photon spectrum of \ver, obtained during
VERITAS observations between MJD 55126 (2009 Oct 22) and MJD 55212 (2010 Jan 16). 
Upper limits at 95\% confidence level are shown for the two highest energy points, 
where the significance is $<2\sigma$.\label{spectrum}}
 \end{figure}

VERITAS data analysis follows the procedure outlined in
\citet{VERITAS_LSI}. 
Cherenkov light from air showers initiated by gamma rays and
cosmic rays triggers the readout of PMT
signals, which are then calibrated and used to reconstruct an image of
the shower in the focal plane. Individual telescope images are
parameterized by simple moment analysis \citep{Hillas85}, and
geometrical reconstruction is used to calculate the arrival direction
of the primary, $\theta$, defined with respect to the position of the
candidate source location on the sky.
The image shape in each telescope is compared with the expected shapes
for gamma-ray showers generated using Monte Carlo simulations and the
differences, averaged over all telescopes, are used to derive the
\textit{mean-reduced-scaled width} and \textit{mean-reduced-scaled
length} parameters, as defined in \citet{VERITAS_LSI}. Gamma-ray-like
events are selected by applying cuts on the various image parameters. 
The cuts used in this work are: \textit{mean-reduced-scaled
width/length} between -1.2 and 0.5, $\theta<0.1^{\circ}$, and at least
three telescope images with an integrated signal size per image equivalent to 
$>94\U{photoelectrons}$. These cuts are designed to provide optimum
sensitivity to a moderately strong point-like gamma-ray source (0.05\U{Crab}) with a Crab-like
differential power-law
photon index of $\sim 2.5$. Over 99.9\% of the background cosmic ray
events are removed by these cuts. The number of background
events remaining in the signal region 
is estimated from 10
off-source regions within the same field of view using the \textit{reflected
region} technique \citep{Aharonian01,refl}.

After data selection, reduction, and signal extraction, an excess of 221 candidate gamma-ray events
over a background of 119 is detected at the location of \rgb, corresponding to a
significance of 15.6 standard deviations according to Eq.~17 in 
\citet[][with $N_{\mathrm{on}}=340$, $N_{\mathrm{off}}=1518$, $\alpha=0.08$]{lima}.
The angular distribution of the excess events is compatible to that of a point-like source given
the instrumental point spread function of $6^\prime$ for 68\% containment radius. The position of the source is derived
by fitting a symmetric 2-dimensional Gaussian to the uncorrelated excess map, resulting in a 
best-fit centroid at 
R.A. = $05^{\rm h}21^{\rm m}45^{\rm s}$,
Decl. = $+21^{\circ}12^{\prime}51.\!^{\prime\prime}4$, 
with a statistical uncertainty of $14^{\prime\prime}$ and a systematic uncertainty of
$25^{\prime\prime}$,
dominated by the telescopes' pointing accuracy. The new TeV source is cataloged as \ver, based on 
the name first reported in \citet{atel-1}.

Figure~\ref{spectrum} shows the time-averaged energy spectrum of \ver, extending from a threshold
energy of
0.2 to $\sim 1\U{TeV}$. It is well described by a power law $dN/dE = N_0
(E/0.4\U{TeV})^{-\Gamma}$ with normalization  $N_0=(1.99\pm0.18_{\mathrm{stat}} \pm
0.80_{\mathrm{syst}})\times 10^{-11} \U{cm^{-2}s^{-1}TeV^{-1}}$ and photon index $\Gamma = 3.44 \pm
0.20_{\mathrm{stat}} \pm 0.30_{\mathrm{syst}}$. The time-averaged integral photon flux
is $F_{>0.2\U{TeV}} = (1.93 \pm 0.13_{\mathrm{stat}} \pm 0.78_{\mathrm{syst}}) \times
10^{-11}\U{cm^{-2}s^{-1}}$, corresponding to $0.092 \pm 0.006\U{Crab}$ \citep[$1\U{Crab}=2.1\times
10^{-10}\UU{cm}{-2}\UU{s}{-1}$,][]{hillas-crab}.
Compatible results were obtained using two independent analysis packages.

\begin{figure}[tdp]
\center
\includegraphics[width=0.99\columnwidth]{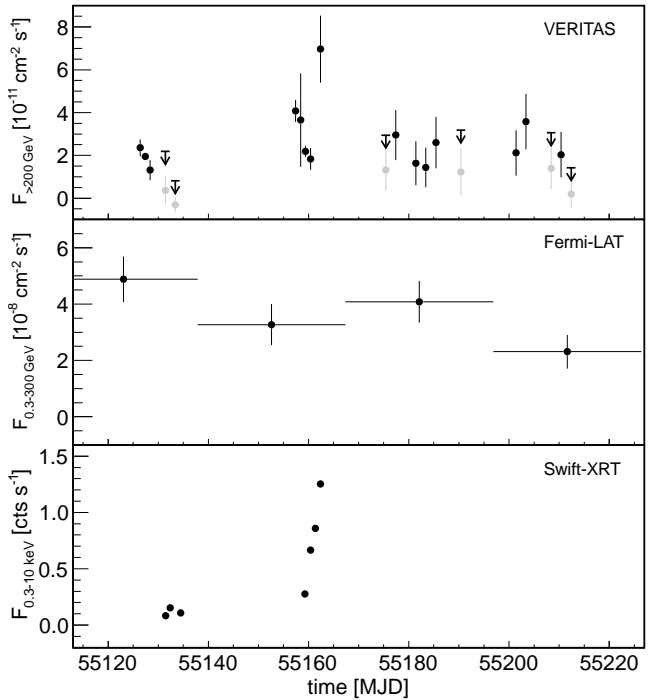} 
\caption{VERITAS ($E>0.2\U{TeV}$), \lat\ ($0.1-300\U{GeV}$), and {\it Swift}-XRT (0.3-10\,keV) light
curves of \ver. Photon fluxes are calculated in 1-day bins for VERITAS and {\it Swift}-XRT,
and 29.5 days for {\it Fermi}-LAT. The VERITAS light curve shows significant flux points (black
dots) when the signal exceeds $2\sigma$, and 95\% confidence level upper limits (black arrows) 
together with flux points
(gray dots) for marginal detections. The error bars on the {\it Swift}-XRT rates are at the 
$\sim 2-7\%$ level, and are not visible in the plot. 
\label{lightcurve}}
\end{figure}

Flux variability was explored by producing a 1-day binned light curve during the VERITAS
observations (Figure~\ref{lightcurve}, {\it top}). After a first detection in 2009 October 22 to 24
\citep{atel-1} with a derived TeV flux of $0.09 \pm 0.01 \U{Crab}$, a later round of
observations revealed a higher flux level, peaking at $0.33 \pm 0.07 \U{Crab}$ on 2009 November 27 
(MJD 55162, hereafter ``TeV flare"). A $\chi^2$ fit to the nightly flux points for constant emission gives a probability of 
$7 \times 10^{-7}$, indicating flux variability at a confidence level of 5.8 standard deviations.

Spectral variability was tested by deriving a ``low-state'' spectrum from the data taken during the first 
period of VERITAS observations (2009 Oct $22-30$, MJD $55126-55134$), and a ``flare'' spectrum from 
the 
night of 2009 November 27. 
A power-law
fit to each spectrum yields $\Gamma_{\mathrm{low}} = 2.92 \pm 0.34$ and $\Gamma_{\mathrm{flare}} =
3.25 \pm 0.72$. The  difference in the reconstructed photon index does not constitute significant
evidence for spectral variability in the TeV band.

\section{{\it Fermi}-LAT observations}

\lat\ is a pair-conversion telescope sensitive to gamma rays in the range from 0.02 to more
than 300\U{GeV} with a field of view of $\sim 2.4\U{sr}$. The effective area of the LAT changes 
with energy and incidence angle, being $\sim 0.8\UU{m}{2}$ for on-axis photons with $E > 
10\U{GeV}$.
Full details about the instrument and its performance are given in \citet{atwood} and 
\citet{on-orbit}.

A GeV source spatially associated with \ver\  was  listed in the LAT 11-month catalog
\citep[1FGL~J0521.7+2114, ][]{1fgl} and confirmed in the second source catalog 
\citep[2FGL~J0521.7+2113, ][]{2fgl} and in the catalog of sources detected above 10\U{GeV} 
\citep[1FHL~J0521.7+2113, ][]{1fhl}.
The region was analyzed in more detail by selecting \textit{source}-class\footnote{ See
{http://fermi.gsfc.nasa.gov/ssc/data/analysis/documentation/
Cicerone/Cicerone\_Data/LAT\_DP.html
} } events with
reconstructed energy $0.3 < E < 300\U{GeV}$ collected during the period of VERITAS observations (MJD
$55126-55212$) in a $20^{\circ}\times 20^{\circ}$ region of interest (RoI) centered at the location
of
\ver. Spectral parameters were extracted by fitting a model containing the Galactic and isotropic
diffuse background\footnote{\texttt{gal\_2yearp7v6\_v0.fits} and
\texttt{iso\_p7v6source.txt}, respectively.} and point sources from the 2FGL catalog. The analysis
was done using binned likelihood as implemented in the \lat\ \texttt{Science Tools v09-26-00} with P7V6
response functions and the standard quality selection cuts described in \citet{2fgl}.
The energy spectrum (light curve) were derived following the procedure detailed in \citet{2fgl}, dividing the data into bins of energy (time) and performing a likelihood analysis with the spectral shape parameters frozen for all sources in the model.

During the period of VERITAS observations, \ver\ was detected by the LAT with a test
statistic of $237$ ($\sim 15 \sigma$). The
derived integral photon flux is $F_{0.3-300\U{GeV}}=(2.8 \pm 0.4)\times 10^{-8}
\UU{cm}{-2}\UU{s}{-1}$,
with a spectrum well-described by a power law with photon index $\Gamma = 1.72 \pm 0.09$. An
identical analysis of an extended dataset collected by {\it Fermi} ($\sim 46$ months) reveals that
\ver\ was in a harder spectral state during the VERITAS observations, compared to a time-averaged
photon index of $\Gamma = 1.97 \pm 0.03$, while the flux level was comparable over both periods.

Variability was tested by extracting the source flux in 29.5-day bins\footnote{The length of the 
bins is equal to the lunar period. Since VERITAS does not observe during full moon, this ensures 
that each epoch of VERITAS observations falls 
within one bin of the LAT light curve.}.
Applying a likelihood-based 
variability test \citep{2fgl} to the 46 months of 
data gives a probability of a constant flux 
of $2 \times 10^{-20}$, indicating 
flux variability in the GeV band at a confidence level of 9.3 standard deviations. 
The 46-month light curve (not shown) exhibits flux changes by a factor of $\sim  4$ between the lowest and highest emission levels.

An additional test for variability above 1\U{GeV} was performed using a Bayesian
Block method \citep{scargle}. 
For this test, \textit{source}-class events with $E > 1\U{GeV}$ were extracted from a $1^\circ$
radius RoI centered on the source coordinates. 
The data were divided into time blocks over which the event rate
was compatible with a constant value. 
The optimal width of the blocks was determined by maximum likelihood analysis using the algorithm described in \citet{jackson}. 
The effective exposure associated to each event is taken into account, correcting for exposure variations caused by the motion of the spacecraft.
A 1\% false-positive threshold was used for detecting variability. The method identifies three 
periods of different constant flux with durations between 310 and 451 days, indicating variability 
above 1\U{GeV}. 
The highest flux state, with boundaries MJD $54993-55419$, includes the VERITAS
observations. However, no significant evidence for
shorter time variability is found in the \lat\ data above 1\U{GeV} during VERITAS observations, or
in coincidence with the X-ray and TeV flare on MJD 55162.

\section{{\it Swift} observations}

After the discovery of TeV emission by VERITAS \citep{atel-1}, {\it Swift}-XRT \citep{Gehrels04}
observations were triggered.
Seven exposures were obtained between 2009 October 27 
and November 27. The total observation time is 16.6\,ks distributed in exposures of 
$\sim 2.5\U{ks}$.

{\it Swift}-XRT data were
analyzed with \texttt{HEAsoft 6.9} and \texttt{XSPEC 12.6.0} using the most recent calibration
files as described in \citet{burrows}. All data were taken in photon counting mode. Pile-up effects
were accounted for by extracting the signal from an annular source region when rates exceeded $0.5
\U{cts\,s^{-1}}$. 

\begin{figure}[tb]
\center
\includegraphics[width=0.99\columnwidth,trim=0cm 0cm 80pt 0pt]{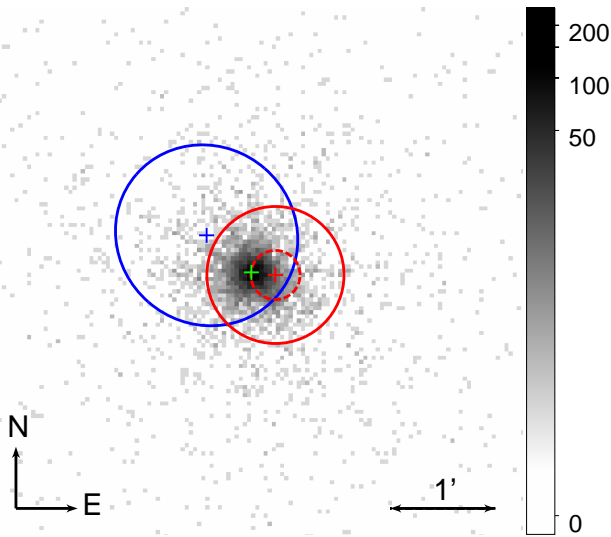}
\caption{ {\it Swift}-XRT cumulative counts map in the $0.3-10\U{keV}$ band, zoomed-in at
the location of \ver. The field is $5^\prime \times 5^\prime$, with gray code indicating counts on
a logarithmic scale. The green cross indicates the best-fit centroid to the {\it Swift}-XRT data.
The red cross shows the position of \ver\ with red dashed circle indicating the VERITAS 68\%
statistical location error and red-solid line showing statistical and systematic error. The blue
ellipse displays 68\% error contour of 2FGL~J0521.7+2113 as measured
by \lat\ \citep{2fgl}.\label{xrt-map}}
\end{figure}

The {\it Swift}-XRT field of view of $23.6^\prime \times 23.6^\prime$ completely
covers the
VERITAS
error circle. A single X-ray source was detected at 
R.A. = $05^{\rm h}21^{\rm m}45.\!^{\rm s}98$,
Decl. = $+21^{\circ}12^{\prime}52.\!^{\prime\prime}9$ (with $1.9^{\prime\prime}$ location
uncertainty), 
spatially coincident
with \rgb, as
shown in Figure~\ref{xrt-map}. The measured, pile-up-corrected count rate is significantly variable
on a daily timescale (Figure~\ref{lightcurve}), with a flux increase up to a factor of
$\sim 15$ between low and high states.
For spectral reconstruction, each observation was binned and fit with an absorbed power law, with
neutral hydrogen column density taken from \citet{kalberla}. The three exposures from 2009 October
$27 - 30$ showed a lower source count rate ($\sim 0.1\U{cts}\UU{s}{-1}$) and were combined for the
spectral analysis. The derived photon index continuously hardens from $2.47\pm 0.10$ when the
emission is lowest to $2.00\pm 0.04$ on the night of the highest X-ray flux (MJD 55162), when \rgb\
reaches a peak flux of $(3.14\pm0.14)\times 10^{-11}\U{erg}\UU{cm}{-2}\UU{s}{-1}$ in the
$2-10\U{keV}$ band.

\section{Optical observations}

\begin{figure}[tdp]
\center
\includegraphics[width=0.82\columnwidth,angle=270]{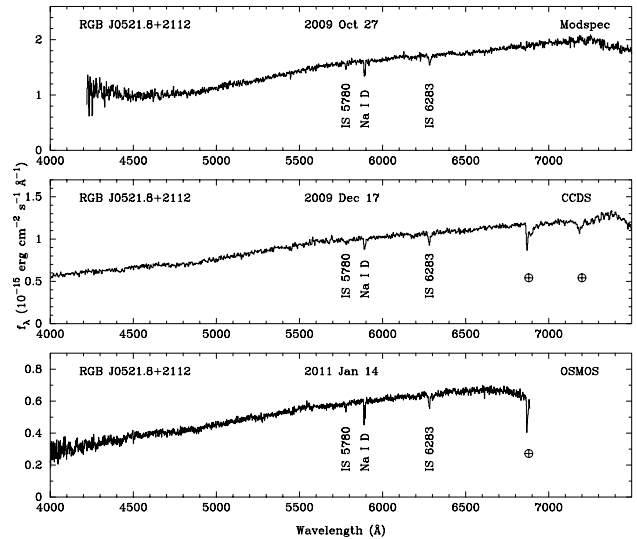} 
\caption{Optical spectra of \rgb\ obtained on the 2.4\,m Hiltner telescope of
the MDM Observatory. The Modspec spectrum was obtained with an exposure of $2\times 600\U{s}$ and a
resolution of $4\U{\AA}$. The exposures for the CCDS and OSMOS spectra were $3\times 1800\U{s}$ with
resolution of $8\U{\AA}$ and $3\U{\AA}$, respectively.
Galactic interstellar absorption lines are labeled
with their wavelength, including the \ion{Na}{1}~D doublet and diffuse interstellar bands at
5780~\AA\ and 6283~\AA.  These are relatively prominent because
of the low Galactic latitude of $-8.7^{\circ}$ and the corresponding
visual extinction $A_V = 1.9$ \citep{sch11} to the source.  Telluric absorption by O$_2$ and H$_2$O
are indicated in the CCDS and OSMOS spectra; in the case of Modspec they
were corrected using the spectrum of a featureless B star.
\label{optspecfig}}
\end{figure}

Beginning with the radio-interferometric position from \citet{bea02}, 
the optical counterpart of \rgb\ was identified 
on the digitized sky survey, with coordinates
R.A. = $05^{\rm h}21^{\rm m}45.^{\rm s}96$,
Decl. = $+21^{\circ}12^{\prime}51.\!^{\prime\prime}6$,
and magnitudes $B2 = 17.32$, $R2 = 15.47$, and $I2 = 15.06$
listed in the USNO B1.0 catalog \citep{mon03}. 
Optical spectroscopic observations were made on three occasions
at the 2.4\,m Hiltner telescope of the MDM Observatory, using three
different CCD spectrographs. 
The optical spectra (Figure~\ref{optspecfig}) are devoid of intrinsic emission or absorption lines
and stellar continuum emission in the wavelength range $4000-7500\U{\AA}$,
classifying the source as a BL Lac-type blazar but not revealing its redshift.
Features marked 
in Figure~\ref{optspecfig} 
are absorption lines from the Galactic interstellar medium at zero redshift.
Recently published observations with the Low Resolution Imaging Spectrograph at the W. M. Keck
Observatory \citep{shaw} show a weak emission feature identified as 
[\ion{N}{2}]$\lambda\lambda\,6548,6583$, which would indicate 
a redshift of $z=0.108$. 
Although this feature cannot be identified in the MDM spectra,
the measurements are not in conflict given the lower level of continuum emission present in the
spectrum by \citet{shaw} and the higher sensitivity of Keck. The first spectrum of \rgb\ in
Figure~\ref{optspecfig} was obtained on 2009 October 27, only 5 days after the first VERITAS
observations.  The calibrated flux
in this spectrum corresponds to $B\sim 17.0$, $R\sim 15.2$, 
slightly brighter than the USNO B1.0 magnitudes.
It appears that the optical flux faded in the subsequent months
and years by $\sim 1$ magnitude, although none of the spectrophotometry
is precise because of the narrow spectrograph slit.

Optical linear polarimetry was performed using the Steward Observatory 1.54\,m Kuiper Telescope,
located on Mt. Bigelow, AZ. \rgb\ was observed on 2011 November 24, 28, and December 1 with the SPOL
CCD
imaging/spectropolarimeter 
\citep{schmidt}.
High signal-to-noise broadband measurements were derived by binning the
polarization spectra in the range of $5000-7000\U{\AA}$.  All measurements are summarized in
Table~\ref{optical} and have been corrected for
statistical bias \citep{wardle}. Given the low Galactic latitude of the source, two field stars were
also observed, 
suggesting a significant interstellar
polarization (ISP) along the sight line to \rgb\ (see Table~\ref{optical}).  The binned 
spectropolarimetry of \rgb, corrected
for an
estimate of the ISP, yields a variable degree of polarization ($P$) between $P=3.74\% \pm 0.15 \%$
with position angle $\theta = 26.8^{\circ} \pm 1.1^{\circ}$ (Nov 24) and
$P=7.26\% \pm 0.16 \%$ at $\theta = 28.2^{\circ} \pm 0.6^{\circ}$ (Dec 1).
Because of the variability in $P$ observed for \rgb,
the object must be intrinsically polarized regardless of the actual level of ISP in this line of sight.

\begin{figure}[tdp]
\center
\includegraphics[width=0.99\columnwidth]{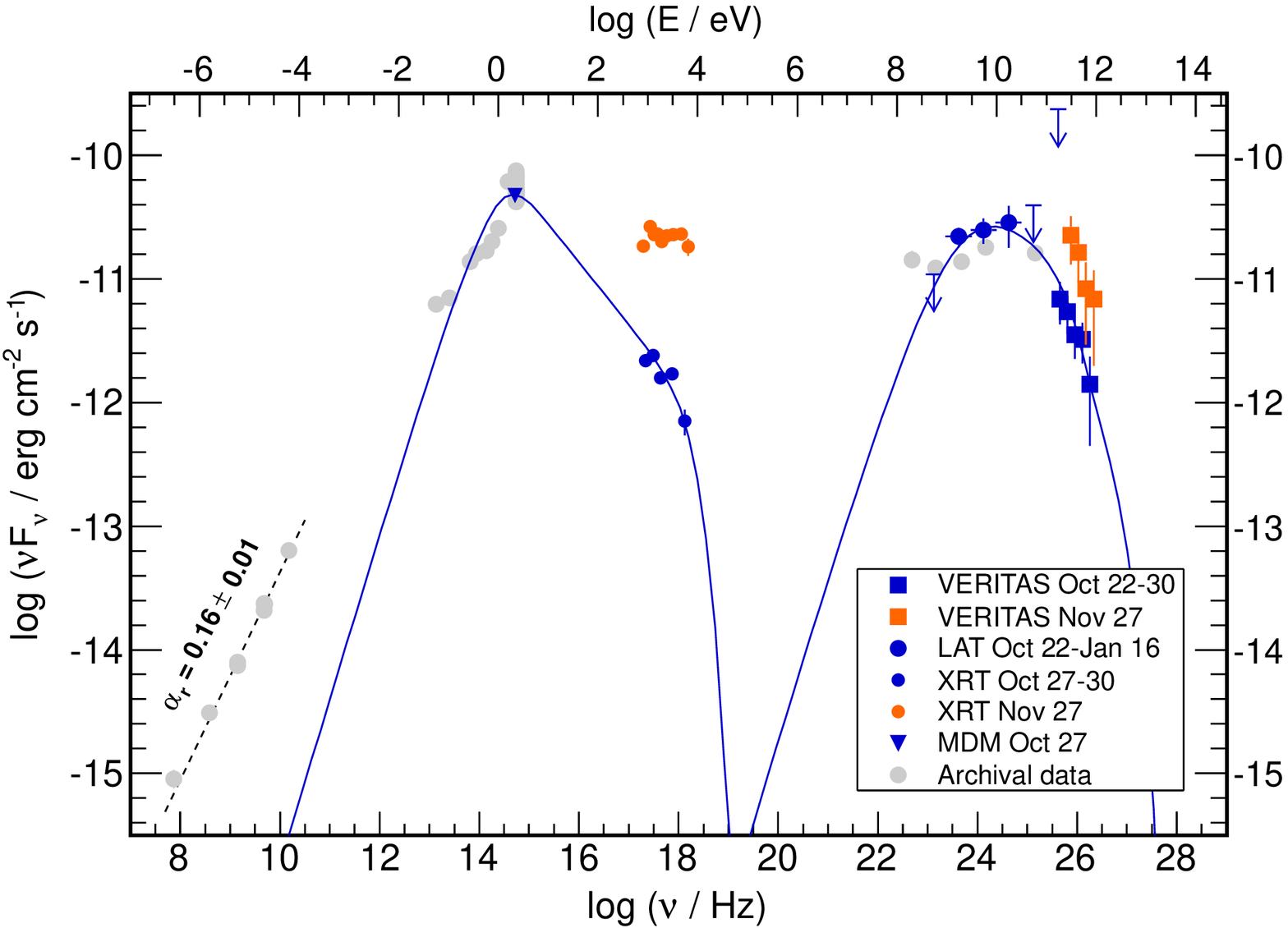} 
\caption{Spectral energy distribution of \ver\ during the VERITAS detection. Optical, X-ray, GeV
and TeV data are shown for the low emission state (blue markers) and for the X-ray and TeV flare
on 2009 November 27 (orange markers). Archival data points are shown in gray: radio 
\citep{cohen,gb6,north20cm,jvaspol,condon}, infrared \citep{wiseptsrc,2mass}, optical
\citep{mon03,catalina}, 
and gamma rays \citep{2fgl}. Archival radio data at 15\U{GHz} is from the OVRO program$^{45}$ 
and was obtained following \citet{ovro}. Optical and infrared data are corrected for Galactic 
extinction using
\citet{sch11}. A dashed black line shows the fit of the radio data to a power law with $F_\nu 
\propto \nu^{-\alpha}$.
The solid blue curve represents a one-zone SSC emission model with parameters adjusted to describe
the low-state data, assuming $z=0.1$. Radio data points are not reproduced by the model, as they 
are expected to sample outer regions of the blazar jet, where the emission becomes optically thin 
to radio waves.
\label{sed}
}
\end{figure}

\begin{table}[b]
\center
\caption{Optical polarimetry of \rgb}
\begin{tabular}{lccc}
\hline\hline
Object  & Date  & $P$ & $\theta$ \\
 &  & [\%] & [$^\circ$] \\
\hline
{\rgb}             & {2011 Nov 24} & $3.24 \pm 0.13$ & $13.6  \pm 1.1$ \\
{\rgb}             & {2011 Nov 28} & $4.47 \pm 0.14$ & $12.7  \pm 0.9$ \\
{\rgb}             & {2011 Dec 01} & $6.50 \pm 0.14$ & $22.0  \pm 0.6$ \\
\hline
{Reference star A} & {2011 Nov 28} & $1.66 \pm 0.07$ & $146.7 \pm 1.2$ \\
{Reference star B} & {2011 Nov 28} & $0.69 \pm 0.04$ & $148.9 \pm 1.7$ \\
\hline
\multicolumn{4}{l}{ISP corrected values}  \\
{\rgb}             & {2011 Nov 24} & $3.74 \pm 0.15$ & $26.8  \pm 1.1$ \\
{\rgb}             & {2011 Nov 28} & $4.83 \pm 0.16$ & $22.7  \pm 0.9$ \\
{\rgb}             & {2011 Dec 01} & $7.26 \pm 0.16$ & $28.2  \pm 0.6$ \\
\hline\hline
\label{optical}
\end{tabular}
\end{table}

\section{Identification of the gamma-ray source}
The observations that led to the discovery of \ver\ were triggered by the identification of a
cluster of photons with $E>30\U{GeV}$ in the first year of \lat\ data (Figure~\ref{Fermi_skymap}).
The most likely counterpart
of the LAT excess was \rgb, which had no classification, although sources in the
ROSAT/Green Bank catalog are generally associated with active galactic nuclei 
\citep[AGN, ][]{rgb}. \rgb\ is the only radio source spatially compatible with the LAT excess in a
high-completeness VLBI sample down to 0.15\U{Jy}, indicative of a radio-loud blazar
\citep[see][]{kovalev}. 
No other spatially associated X-ray, UV, or compact radio
source was found in archival catalogs.

Immediately after the VERITAS detection, X-ray and optical observations were triggered to confirm
the extragalactic nature of \ver. {\it Swift}-XRT establishes \rgb\ as the
only bright X-ray source inside the $39^{\prime\prime}$ VERITAS error circle (Figure~\ref{xrt-map}).
The
identification of \ver\ with \rgb\ 
is further supported by the 
detection of a TeV flare 
simultaneously seen in X-rays (Figure~\ref{lightcurve}). 
However, the statistical significance of the overall
X-ray/TeV flux correlation cannot be assessed due to the limited number of flux measurements in both
bands.

The flux variability observed in the TeV band further suggests that \ver\ could indeed be an AGN.
To date, all known variable TeV sources are AGNs,
with the exception of the 
four detected gamma-ray binaries (LSI~+61~303, HESS~J0632+057, PSR~B1259-63, and LS~5039) and the 
Crab pulsar, which is variable at a much shorter timescale (33\U{ms}).

The
optical counterpart of \ver\ was subsequently identified through observations at MDM. Optical
spectroscopy revealed a continuum-dominated spectrum (Figure~\ref{optspecfig}), unambiguously
identifying \ver\ as a BL Lac-type blazar.

\footnotetext[44]{http://www.astro.caltech.edu/ovroblazars}

\begin{table}[tdp]
\center
\caption{Observational properties of \ver\ during VERITAS observations \label{properties}}
\begin{tabular}{lcc}
\hline\hline
& low state & flare \\
\hline
Radio spectral index ($\alpha_\mathrm{r}$)
\footnote{$F_\nu \propto \nu^{-\alpha_{\mathrm{r}}}$} 
& \multicolumn{2}{c}{$0.16 \pm 0.01$}  \\
Optical polarization ($P$) & \multicolumn{2}{c}{$3.7\%-7.3\%$} \\
X-ray flux ($F_{2-10\mathrm{keV}}$)
\footnote{$10^{-11}\U{erg}\UU{cm}{-2}\UU{s}{-1}$} 
&$0.16\pm0.2$ & $3.14\pm0.14$ \\
X-ray photon index ($\Gamma_\mathrm{x}$)
\footnote{$dN/dE \propto E^{-\Gamma}$}
& $2.5 \pm 0.1$ & $2.0 \pm 0.1$\\
GeV photon flux ($F_{0.3-300\mathrm{GeV}}$)
\footnote{$10^{-8}\UU{cm}{-2}\UU{s}{-1}$}
& \multicolumn{2}{c}{$2.8\pm0.4$} \\
GeV photon index ($\Gamma_\mathrm{GeV}$) 
& \multicolumn{2}{c}{$1.7\pm0.1$} \\
TeV photon flux ($F_{>0.2\mathrm{TeV}}$)
\footnote{Crab \citep[$2.1\times 10^{-10}\UU{cm}{-2}\UU{s}{-1}$,][]{hillas-crab}}
& $0.09 \pm 0.01$ & $0.33 \pm 0.07$ \\
TeV photon index ($\Gamma_\mathrm{TeV}$) & $2.9 \pm 0.3$& $3.3\pm0.7$\\ 
TeV luminosity ($L_{>0.2\mathrm{TeV}}$) 
\footnote{$\U{erg}\UU{s}{-1}$}
& $2.4\times10^{44}$ & $8.8\times10^{44}$\\
\\
Radio loudness ($R_{\mathrm{rB}}$)
\footnote{$R_{\mathrm{rB}}=L_{5\mathrm{GHz}}/L_{\mathrm{B_{mag}}}$}
& \multicolumn{2}{c}{69}  \\
Radio-to-optical slope ($\alpha_{\mathrm{ro}}$)
\footnote{$\alpha_{\mathrm{ro}} =-\log(F_{1.4\mathrm{GHz}}/F_{\mathrm{R_{mag}}}) /
\log(\nu_{1.4\mathrm{GHz}}/\nu_{\mathrm{R_{mag}}})$} 
& 0.42 & 0.46\\
Optical-to-X-ray slope ($\alpha_{\mathrm{ox}}$)
\footnote{$\alpha_{\mathrm{ox}} =-\log(F_{\mathrm{R_{mag}}}/F_{1\mathrm{keV}}) /
\log(\nu_{\mathrm{R_{mag}}}/\nu_{1\mathrm{keV}})$} 
& 1.31 & 0.77\\
Compton dominance ($R_C$)
\footnote{$R_C=L_{\mathrm{HE}}/L_{\mathrm{sy}}$}
& $1.8$ & $0.8$\\
Redshift ($z$)
& \multicolumn{2}{c}{$0.108$\,\footnote{\citet{shaw}}\ ($<0.34$)\footnote{derived in
Section~\ref{z-limit}}} \\
\hline\hline
\end{tabular}
\end{table}

\section{Observational properties of \ver}

The main observational properties of \ver\ are summarized in
Table~\ref{properties}.
Taking the 5\U{GHz} flux from \citet{north20cm} and the $B$-band apparent magnitude
from MDM observations, the radio-to-optical luminosity ratio for \ver\ is
$R_{\mathrm{rB}}=L_{5\mathrm{GHz}}/L_{\mathrm{B_{mag}}}=
69$, implying a radio-loud AGN 
\citep[$R_{\mathrm{rB}}>10$, ][]{kellermann} with prominent radio jet or lobe emission
\citep{urry}. 
The spectrum
of the unresolved core emission of \ver\ in the radio band 
is well described by a power law, with $F_\nu
\propto \nu^{-\alpha_{\mathrm{r}}}$ and spectral index
$\alpha_{\mathrm{r}}=0.16 \pm 0.01$ (Figure~\ref{sed}),
compatible with $\alpha_{\mathrm{r}} \leq 0.5$, 
characteristic of jet-dominated flat-spectrum radio sources. 

\begin{figure}[tdp]
\center
\includegraphics[width=0.95\columnwidth,angle=270,clip,trim=1.55cm 0cm 0cm
0pt]{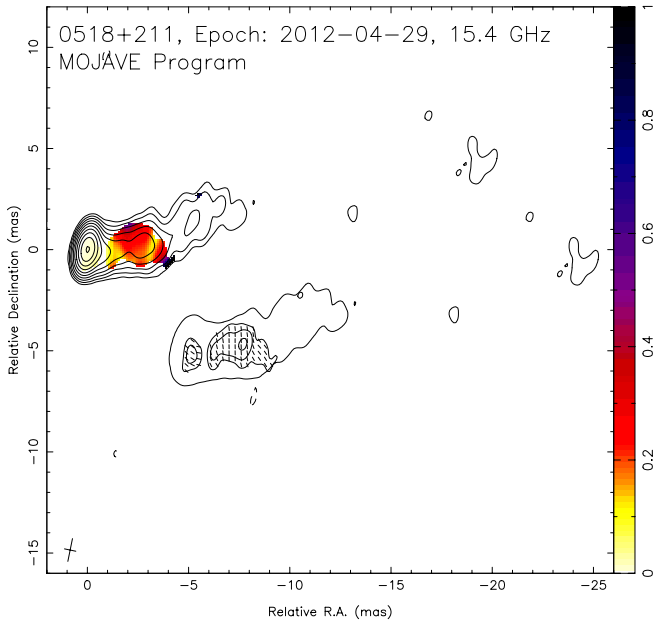} 
\caption{15 GHz MOJAVE VLBA composite image of \rgb\ on 2012 April 29. The top left image shows 
contours of total intensity, with fractional linear polarization overlaid in color scale. The
bottom right image shows 
contours of linear polarization, plus a single outermost contour of total intensity. 
Electric polarization vector directions 
are plotted as equal-length ticks. The contour levels for total intensity are 0.5 mJy/beam in steps
of 2, while those for polarized intensity are for 0.7 mJy/beam in steps of 2. The restoring beam is
indicated in the bottom left corner, and has Gaussian FWHM dimensions 1.15 by 0.57 mas, with major
axis at -11.4 degrees from north.
\label{radio}}
\end{figure}

The radio jet of \rgb\ has been imaged on milliarcsecond scales
in 5 epochs between 2009 October and 2012 April with the VLBA at 15 GHz as
part of the MOJAVE program \citep{mojave}. The
radio morphology consists of a bright radio core and an apparent
one-sided jet that extends for $\sim 20\U{mas}$ to the west while curving
slightly to the northwest (Figure~\ref{radio}). The radio core is very compact, with a
brightness temperature above $10^{11}\U{K}$ at all epochs
according to Gaussian model fits. The model fits to individual
features in the jet do not reveal significant proper motions over the 2.5-year observation interval.
During this time, however, there were
significant changes in the polarization of the core and the inner $1.5\U{mas}$ 
of the jet. The core remained weakly linearly polarized ($< 1\%$),
but increased steadily in polarized flux density from 0.7 to 1.6 mJy. 
The downstream jet polarization was typically much higher  (up to $25\%$) with electric
polarization 
vectors perpendicular to the jet ridgeline, 
consistent with optically thin synchrotron emission associated with  a 
relatively well-ordered, longitudinal jet magnetic field.
A bright feature $\sim 1\U{mas}$
downstream of the core briefly flared in total intensity and
polarization sometime between 2010 March and 2010 October, but
returned to its earlier level by 2011 July. These short-timescale
changes, as well as the one-sided jet morphology, high core
compactness and jet polarization are very similar to those seen in
highly Doppler-boosted blazar jets in the MOJAVE sample \citep{lister04,lister09}.

Infrared observations of \ver\ in the WISE Source
Catalog\footnote[45]{\url{http://wise2.ipac.caltech.edu/docs/release/allsky}} show magnitudes of
10.6, 9.8, 7.6, and 5.6 in the 3.4, 4.6, 12 and $22\U{\mu m}$ bands, respectively.
Its infrared colors are similar
to those of known gamma-ray blazars \citep{massaro}, providing further support for its
identification as a TeV blazar \citep{atel-wise}.

The optical polarimetry reveals a highly polarized object that shows significant variations in both
the level and position angle of polarization on time scales shorter than a week.  This is strong
evidence that a good fraction of the optical continuum of \rgb\ is produced by
synchrotron radiation from regions with well-ordered magnetic fields. During the
epoch of the Steward Observatory measurements, the position angle of the polarization
($20-30^\circ$) was intermediate between being parallel and perpendicular to the axis of the radio
jet.

\begin{figure}[tdp]
\center
\includegraphics[width=0.999\columnwidth]{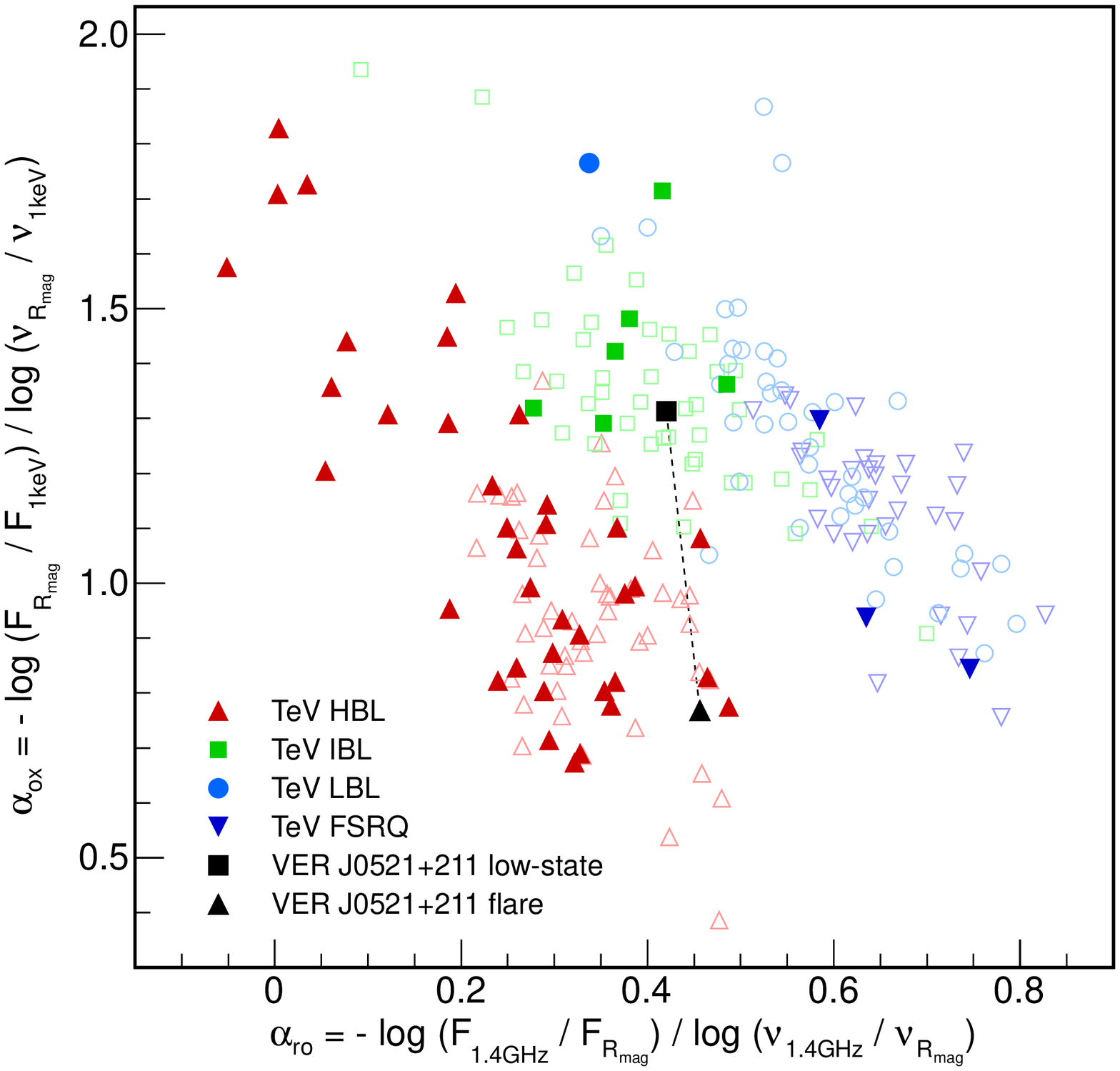} 
\caption{ Spectral slopes of the synchrotron component for gamma-ray blazars.
The effective spectral indices $\alpha_{\mathrm{ro}}$ and $\alpha_{\mathrm{ox}}$ are defined in the
usual way between 1.4\U{GHz}, 6590\U{\AA} and 1\U{keV}. SED classifications for the 47 known TeV
blazars (filled markers) are taken from TeVCat \citep{tevcat}.
Multiband fluxes are obtained from \citet{bzcat} with the exceptions of RX~J0648.7+1516,
HESS~J1943+213 and MAGIC~J2001+435 \citep{condon,mon03,rass,bassani}. Empty markers show
GeV-detected blazars from \citet{2fgl} present in \citet{bzcat}. 
\ver\ is pictured in a ``low-state'' (2009 Oct $22-30$) showing a spectral shape
characteristic of IBLs, and in ``flare'' (2009 Nov 27), when it shows HBL-like properties.
\label{synch}}
\end{figure}

BL Lac-type blazars are usually classified 
as low, intermediate, or high-frequency peaked BL Lacs 
according to the estimated peak frequency of their
synchrotron component \citep[LBL, $\log(\nu_{\mathrm{sy}} / \mathrm{Hz}) < 14$; IBL, $14-15$; 
HBL, $>15$; e.g.][]{1lac}.
Figure~\ref{synch} shows the effective spectral indices between radio, optical, and X-ray
bands for all known TeV blazars.
Simultaneous flux measurements during 
``low-state'' (2009 Oct $22-30$) suggest that \ver\
has synchrotron properties similar to those of the known IBLs. 
However, during the X-ray and TeV flare, 
the brightening and hardening of the
X-ray emission causes 
the synchrotron component of \ver\ to show HBL-like properties
(Figure~\ref{synch}). 
Spectral changes in the synchrotron component are not uncommon and have been reported in other TeV
blazars \citep{massaro08,bllac}.

\ver\ also shows similar luminosity in the synchrotron and high-energy peak of the SED
(Figure~\ref{sed}), implying a Compton
dominance ($R_C=L_{\mathrm{HE}}/L_{\mathrm{sy}}$) in the range $0.8\lesssim R_C \lesssim
1.8$. Such values are similar to those of high-power blazars \citep{meyer12}, which show synchrotron
peak frequencies lower than ``classical'' HBLs \citep[$\log(\nu_{\mathrm{sy}} / \mathrm{Hz})
\lesssim 14.5$,][]{meyer11}.

With a low-state TeV flux of $\sim 0.1\U{Crab}$ measured over three months of observations, and a
flare
of $\sim 0.3\U{Crab}$, \ver\ is one of the brightest TeV blazars newly discovered by the current
generation of Cherenkov telescopes.
Other blazars, such as 3C~279 \citep{279},
W~Comae \citep{wcom}, or
PKS~1222+216 \citep{1222}, have been detected at flux levels higher than $0.1\U{Crab}$ during short
periods of time ($\leq 1\U{week}$) and then quickly fall below instrument sensitivity.
Despite being a relatively bright TeV source, \ver\ was never
identified as a TeV-blazar candidate before the VERITAS detection. 
Given its low Galactic latitude of $-8.7^{\circ}$, \ver\ was poorly characterized in the optical
and X-ray bands and was not included in blazar catalogs, which are usually
incomplete at low Galactic latitudes or avoid the Galactic plane altogether
\citep[e.g.,][]{cgrabs,bzcat}, as diffuse radio emission, confusion with local
radio sources, and heavy optical extinction make candidate blazars difficult to identify.
Therefore, TeV-candidate catalogs \citep[e.g.,][]{costamante,donato} lacked sufficient
multiwavelength information on \ver\ and a firm identification as a blazar, required to predict its 
TeV flux.

\section{Redshift upper limit}
\label{z-limit}

An upper limit on the redshift of \ver\ can be derived from its TeV spectrum.
Following the approach in \citet{mazin} \citep[see
also][]{hess-ebl,mazin-raue}, the intrinsic 
TeV spectrum of \ver\ is reconstructed from the VERITAS measured spectrum (Figure~\ref{spectrum}) by correcting 
for EBL absorption, assuming the density model of \citet{dominguez}, under varying assumptions of
$z$. The intrinsic 
spectrum is then fit with a power law of the form $dN/dE \propto E^{-\Gamma^{\star}}$. Larger values of $z$ result in 
a harder reconstructed photon index $\Gamma^{\star}$. Classical leptonic emission models predict
$\Gamma^{\star}>1.5$ \citep[see discussion in ][]{hess-ebl}. Under this assumption, a redshift upper
limit of $z<0.34$ is derived for \ver\ at 95\% confidence level. 
An even more conservative 
redshift upper limit can be obtained by allowing an intrinsic spectral index as hard as 
$\Gamma^{\star} \sim 0.7$, as 
suggested in  \citet{kata} \citep[see also][]{aha,sitarek}. Under this less restrictive assumption of 
$\Gamma^{\star}>0.7$, the redshift of \ver\ is constrained to $z < 0.44$.

The redshift upper limits derived from the TeV spectrum of \ver\ are in agreement with the recent
measurement of $z=0.108$ based
on a single optical spectral feature \citep{shaw}.

\begin{table}[b]
\center
\caption{SED modeling parameters\label{model}}
\begin{tabular}{lrcc}
\hline\hline
Parameter & Symbol & Value \\
\hline
Electron distribution &&\\
Electron power &$L_e$ [erg s$^{-1}$]&  $7.7\times10^{44}$    \\
Low-energy cutoff &$\gamma_{min}$  &  $3.5\times10^{4}$     \\
High-energy cutoff &$\gamma_{max}$  &  $2.0\times10^{6}$   \\
Injection index &$q_e$                 &      3.0          \\ \\
Blob radius &$R_b$ [cm]           &     $4.0\times10^{17}$      \\
Magnetic field &$B$ [G]               &   0.0025      \\
Bulk Lorentz factor &$\Gamma$            &    30       \\
Escape parameter &$\eta_{esc}$           &    300      \\
Redshift (assumed) &$z$                          &   0.10    \\
\hline
\hline
\end{tabular}
\end{table}

\section{Spectral energy distribution}

After a successful identification of \ver\ as a new TeV blazar, the available multiwavelength
data were combined to construct a spectral energy distribution (SED), shown in Figure~\ref{sed}.
Most commonly accepted models attribute the low-energy emission
component to synchrotron radiation by relativistic electrons in the jet magnetic field, and the
high-energy to inverse-Compton scattering of ambient photons off the same
electron population \citep[see, e.g.,][]{maraschi}.

The multiwavelength SED can be described with a one-zone synchrotron self-Compton (SSC) emission
model as described in \citet{bottcher}. The model parameters are adjusted
to describe the quasi-simultaneous spectral points obtained during the ``low'' emission state (2009
Oct $22-30$). 
Models were tested assuming $z=0.05, \,0.1, \,0.15, \,0.2, \,0.25$, although only results for
$z=0.1$ are discussed, being the assumption that most closely matches the tentative redshift of $z=0.108$.

The model parameters describing the low-state SED of \ver\ are listed in Table~\ref{model}, and 
show a slightly low magnetic field and large emitting region compared to other TeV blazars 
\citep{heike}.
One-zone SSC models can only describe Compton-dominated systems ($R_C \gtrsim 1$) like \ver\ with
very low magnetic fields, resulting in a strongly particle-dominated jet ($L_B/L_e < 0.01$, 
$L_B$ and $L_e$ being the magnetic and particle power in the jet, respectively). External Compton
models \citep[EC,][]{dermer} add a second population of low-energy photons where relativistic 
electrons inverse-Compton scatter, increasing the level of high-energy emission, with jet
energetics close to equipartition ($L_B/L_e \sim 1$). 
However, given the scarcity of simultaneous observations,  particularly  in  the synchrotron
component, an EC model applied to the SED of \ver\ would be severely
underconstrained.

\section{Summary and conclusions}

VERITAS detected a new TeV source: \ver, spatially associated
with the radio and X-ray source \rgb. 
Follow-up observations in the optical and X-ray bands unambiguously identify \ver\ as a new blazar
of
the BL~Lac type, displaying all the defining properties of blazars in radio, infrared, and optical 
wavelengths.
The detected TeV emission is variable on daily timescales with
an integral flux of
$\sim0.09-0.33\U{Crab}$ measured between 0.2 and $\sim 1\U{TeV}$, and a time-averaged spectrum 
compatible with
a power law with photon index $\Gamma = 3.44 \pm 0.20_{\mathrm{stat}} \pm 0.30_{\mathrm{syst}}$. 
During the observing campaign that covered 2009 October to 2010 January, \ver\ transitioned to a
high state on 2009 November 
27, when the nightly flux increased by a factor of $\sim 3$ in the TeV range and by a factor of
$\sim 15$ in X-rays, 
compared to the observed baseline values. 
X-ray observations show a trend of spectral hardening with increasing flux, while no significant
spectral variability was 
found at TeV energies.

Observations from radio to X-ray frequencies show a multitude of evidence for a
synchrotron origin of the emission from \ver\ below a few keV, as expected for a blazar: a flat
spectrum in the radio band with a turnover towards 
infrared frequencies, polarized emission in the optical band and in the 15\U{GHz} radio images from 
VLBA,
and a nonthermal power-law spectrum in X-rays.
The radio images also
show electric polarization vectors perpendicularly aligned to the jet ridgeline, suggesting a
relatively well-ordered
magnetic field in the direction of the jet axis.
The multiband spectral shape of the synchrotron component of \ver\ is similar to that of known TeV IBLs. However, 
during the TeV and X-ray flare its synchrotron properties are closer to those of HBLs. Optical
spectrometry with MDM could not be used to derive a redshift, although recent measurements suggest
$z=0.108$. The TeV spectrum of \ver\ constrains its redshift to $z<0.34$ 
under the assumption that the intrinsic TeV photon index of the source is $\Gamma^\star > 1.5$,
which is the limit obtained for standard leptonic emission models. 

The high-energy emission from \ver\ peaks in the gamma-ray band, between 10 and 200\U{GeV}, and can be 
described with a leptonic one-zone SSC emission model. The resulting model parameters would indicate 
a relatively weak magnetic field of $\lesssim 0.01\U{G}$ and a particle-dominated jet. Similar 
objects have been modeled by  adding an EC component to the SSC emission \citep{wcom, 3c66a}, or 
considering a structured jet with a  fast-moving spine and a slower outer layer \citep{0716}. These 
models could presumably describe the SED of \ver, and generally reach solutions closer to 
equipartition.  However, being more complex than SSC emission, SSC+EC and structured jet models 
have more free parameters, and would be underconstrained given the data available for \ver. 

Because of its low Galactic latitude, \ver\ observations were not triggered by its X-ray
properties like most TeV-candidate blazars, but by a cluster of $E>30\U{GeV}$ photons detected in
the \lat\ public data released after the first year of observations. 
Selection criteria based on \lat\ data have been successful in identifying other new TeV blazars
\citep{1222,0413,0447}, particularly at low Galactic latitudes where selections based on radio and
X-ray data are less powerful due to Galactic extinction. 
The detection of \ver\ adds to previous results that have demonstrated the strength of GeV-band
\citep[e.g.,][]{muk,vanden,kara} and TeV-band observations 
\citep[e.g.,][]{j2001,0648,hess-1943}
as a tool to identify blazars
located behind the Galactic plane.

With a TeV flux between 0.09 and 0.33 Crab, \ver\ ranks among the brightest known TeV blazars, and
can be detected with current ground-based Cherenkov telescopes in less than one hour exposure. 
Assuming a redshift of 0.108, the TeV luminosity of \ver\ is $L_{>0.2\mathrm{TeV}} \sim 2.4 \times 
10^{44}\U{erg}\UU{s}{-1}$, larger than that of the ``classical'' northern TeV blazars
(Mrk~421, Mrk~501, and 1ES~1959+650), which sample the low luminosity end of the population of TeV
blazars \citep{obs-tev}.
Given the observed variability and its bright TeV flux, future
multiwavelength observations of \ver\ will be able to extend the detailed time-resolved spectral modeling
available for nearby HBLs \citep{sed-1959,sed-421,sed-501} to a more luminous non-HBL blazar.

\acknowledgments
\begin{small}

VERITAS is supported by grants from the U.S. Department of Energy Office of Science, the U.S.
National Science Foundation and the Smithsonian Institution, by NSERC in Canada, by Science
Foundation Ireland (SFI 10/RFP/AST2748) and by STFC in the U.K. We acknowledge the excellent work of
the technical support staff at the Fred Lawrence Whipple Observatory and at the collaborating
institutions in the construction and operation of the instrument.

The \textit{Fermi}-LAT Collaboration acknowledges support from a
number of agencies and institutes for both development and the
operation of the LAT as well as scientific data analysis. These
include NASA and DOE in the United States, CEA/Irfu and IN2P3/CNRS in
France, ASI and INFN in Italy, MEXT, KEK, and JAXA in Japan, and the
K. A. Wallenberg Foundation, the Swedish Research Council, and the
National Space Board in Sweden. Additional support from INAF in Italy
and CNES in France for science analysis during the operations phase is
also gratefully acknowledged.

ME acknowledges support from the NASA grants NNX10AP66G and NNX12AJ30G.
YYK was supported in part by the Russian Foundation for Basic Research (projects 11-02-00368 and
12-02-33101), the basic research program ``Active processes in galactic and extragalactic objects''
of the Physical Sciences Division of the Russian Academy of Sciences, and the Dynasty Foundation.
MB acknowledges support by the South African Department of Science
and Technology through the National Research Foundation under NRF
SARChI Chair grant no. 64789.

This research has made use of data from the MOJAVE database that is maintained by the MOJAVE team 
\citep{mojave}. The MOJAVE project is supported under NASA-Fermi grants NNX08AV67G  and 11-Fermi11-0019. The authors thank Julie Skinner for obtaining, as a target of opportunity,
the first MDM spectrum of \rgb\ used in this paper; and Talvikki Hovatta for providing the OVRO
radio data.
Observations at Steward Observatory were supported by the NASA Fermi Guest Investigator Program grant NNX09AU10G. Finally, the authors thank the \textit{Swift} team for accepting and carefully scheduling the target of opportunity observations of \ver\ that were used in the paper and for support from 
the {\it Swift} Guest Investigator program, NASA grant NNX10AF89G.

\end{small}

\end{document}